\documentclass[traditabstract]{aa}
\pdfoutput=1
\usepackage{graphicx}
\usepackage{amsmath,amsfonts,amssymb,tabu}
\usepackage{txfonts}
\usepackage[breaklinks,colorlinks,citecolor=blue,pdfa=true]{hyperref}
\usepackage{color}
\usepackage{fixltx2e}
\usepackage{natbib,twoopt}
\usepackage{url}
\usepackage{multirow}
\usepackage{epsf}
\usepackage{epsfig}
\usepackage{longtable}
\usepackage{float}
\usepackage{subfig}
\usepackage{caption}
\usepackage[dvipsnames]{xcolor}

\definecolor{mypink1}{RGB}{219, 48, 122}
\usepackage{ifthen}
\usepackage[T1]{fontenc}
\usepackage{lmodern}
\usepackage{ifxetex,ifluatex}
\usepackage{latexsym}
\usepackage{pdfpages}
\usepackage{dblfloatfix}
\usepackage{morefloats}
\usepackage{caption}
\usepackage{lscape}
\usepackage{mathtools}

\def\tex {\ifmmode{{T}_{\rm ex}}\else{$T_{\rm ex}$}\fi}
\def\tmb {\ifmmode{{T}_{\rm mb}}\else{$T_{\rm mb}$}\fi}
\def\ci     {\ifmmode{{\rm C}{\rm \small I}}\else{C\ts {\scriptsize I}}\fi}
\def\hi     {\ifmmode{{\rm H}{\rm \small I}}\else{H\ts {\scriptsize I}}\fi}
\def\hh     {\ifmmode{{\rm H}_2}\else{H$_2$}\fi}

\def\ts     {\thinspace}
\def\kms    {\ifmmode{{\rm \ts km\ts s}^{-1}}\else{\ts km\ts s$^{-1}$}\fi}
\def\msol   {\ifmmode{{\rm M}_{\odot}}\else{M$_{\odot}$}\fi}
\def\lsol   {\ifmmode{{\rm L}_{\odot}}\else{L$_{\odot}$}\fi}
\def\zsol   {\ifmmode{{\rm Z}_{\odot}}\else{Z$_{\odot}$}\fi}

\begin{document}

\title{SAGAN-II : Molecular gas content of giant radio galaxies
\thanks{Based on observations carried out with IRAM-30m}}

\author{P. Dabhade \inst{1,2}\thanks{E-mail: pratik@strw.leidenuniv.nl}
\and
F. Combes \inst{3,4}
\and
P. Salom\'e \inst{3}
\and
J. Bagchi \inst{2}
\and
M. Mahato \inst{2}
           }
%\offprints{P. Dabhade}
\institute{Leiden Observatory, Leiden University, P.O. Box 9513, NL-2300 RA, Leiden, The Netherlands 
\and %2
Inter-University Centre for Astronomy and Astrophysics (IUCAA), Pune 411007, India
\and %3
Sorbonne Universit\'e, Observatoire de Paris, Universit\'e PSL, CNRS, LERMA, 75014 Paris, France
\and %4
Coll\`ege de France, 11 Place Marcelin Berthelot, 75231 Paris, France
              }

   \date{Received  2020/ Accepted  2020}

   \titlerunning{Molecular gas in GRGs}
   \authorrunning{P. Dabhade et al.}

   \abstract{
Radio galaxies with jets of relativistic particles are usually hosted by massive elliptical galaxies with active nuclei powered by accretion of interstellar matter onto a supermassive black hole. In some rare cases ($<$5\%), their jets drive the overall structure to sizes larger than 700 kpc, and they are called giant radio galaxies (GRGs).
A very small fraction of the population of such radio galaxies contains molecular and atomic gas in the form of rings or discs that can fuel star formation. The origin of this gas is not well known; it has sometimes been associated
with a minor merger with a gas-rich disc galaxy (e.g. Centaurus A) or cooling of material from a hot X-ray atmosphere (e.g. cooling flows). The giant radio jets might be the extreme evolution of these objects, and they can teach us about the radio galaxy evolution. \\ 
We selected 12 targets from a catalogue of 820 GRGs that are likely to be in a gas-accretion and star formation phase. The targets were selected from the mid-infrared to contain heated dust. We report here the results of IRAM-30m observations, the molecular gas content, and the star formation efficiency, and we discuss the origin of the gas and disc morphology. Three out of our 12 targets are detected,  and for the others, we report significant upper limits. We combine our three detections and upper limits with four additional detected GRGs from the literature to discuss the results. Most of the GRG targets belong to the main sequence, and a large fraction are in the passive domain. Their star formation efficiency is comparable to normal galaxies, except for two galaxies that are deficient in molecular gas with a short ($\sim$ 200~Myr) depletion time, and a quiescent gas-rich giant spiral galaxy. In general, the depletion time is much longer than the lifetime of the giant radio jet.
}


\maketitle 

%---------------------------------------------------------------

\section{Introduction}
\label{intro}

It has recently become evident that galaxies and their central supermassive black holes (SMBHs) co-evolve \citep{hbrev14}, and new research continues to bring forth new facts that improve our understanding.
The simultaneous growth of SMBHs and bulges of galaxies might be a consequence of AGN feedback, which plays a fundamental regulating role on star formation (SF), and might result in the observed tight correlation between their masses \citep[e.g.][]{Croton2006, Magorrian1998, Marconi2003}. Energy transfer from the AGN can produce outflows and temporarily starve galaxies of fresh gas, thus quenching the star formation \citep[e.g.][]{Cicone2014}. It is now understood that the galaxy's SMBH can play a vital role in its growth and evolution, and vice versa. 
The most spectacular manifestations of massive black holes in
AGNs are the powerful bipolar relativistic jets that produce twin-lobed structures on $\sim$ 10$^2$-10$^3$ kpc scales. Most of the time, evidence for SF quenching is obtained from studying radio-loud AGN, where molecular gas is entrained by the radio jets. These radio galaxies (RGs) are a unique opportunity to study the AGN-fueling process during the radio-mode phase as well as the AGN feedback.

The AGN feedback and outflows can be studied through the cold molecular gas properties of the galaxy, and  current studies have been restricted to mostly luminous AGNs.
In the past three decades, several studies at lower \citep{Evans2005,TANGO2010,prandoni2010,smolcic11,qsalome15,lanz16,ruffa19} and higher redshifts \citep{hughes97,vanojik97,carlos03,emonts14,nesvadba17} have been conducted to probe the molecular gas content, SFR, and cold gas as potential fuel supply in RGs.
A CO study of a sample of low-luminosity RGs carried out by \citet{prandoni2010} supported the idea that RGs might be fuelled by cold gas, restricted to kiloparsec-scale nuclear discs or rings around the AGN. \citet{ruffa19} have shown a strong possibility that jet-cold gas interacts in a few objects. They also reported inflows inferred from dominant radial motions, indicating that cold gas fuels the RGs.

The members of an extreme sub-class of RGs, which grow to megaparsec scales and are relatively rare, are called the Giant Radio Galaxies (GRGs). The projected linear size of GRGs is greater than 0.7 Mpc and can grow to about 5 Mpc, which is comparable to large galaxy clusters. This makes them one of the largest single astrophysical objects known to date. They have accreting SMBHs of mass $10^{8} - 10^{10}$~$\rm M_{\odot}$,  which  under certain conditions launch collimated bipolar relativistic jets orthogonal to their accretion disc \citep{Lynden-Bell1969,Begelman1979}. Like normal radio galaxies, GRGs are almost always hosted by bright elliptical galaxies.
GRGs were discovered \citep{willis74} almost 20 years after the discovery of RGs \citep{JD53}, and since then, only about 820 \citep{DabhadeSAGAN20} of them have been found, compared to thousands of RGs.

Some CO observations of GRGs have been attempted before (3C236: \citealt{Labiano2013}, 3C 326: \citealt{Nesvadba2010}) but were mostly single-object studies.
\citet{Saripalli2007}, as part of their work on restarted RGs, observed five GRGs with the 15m Swedish-ESO Millimetre Telescope (SEST) in the CO(1-0) and CO(2-1) line transitions and obtained only upper limits for all their GRGs.

To understand these extreme cosmic radio sources better, and specifically, to address the key questions about their size and power, we have initiated a project called Search and Analysis of GRGs with Associated Nuclei (SAGAN\footnote{\url{https://sites.google.com/site/anantasakyatta/sagan}}), whose pilot study was presented in \cite{Dabhade2017}. In this project, we have built  an uniform database of all known GRGs, which was lacking before \citep{DabhadeSAGAN20}. With this GRG database, we aim to carry out multi-wavelength studies of their host galaxies, focusing on physical 
properties such as their black hole accretion rate; the excitation type, which is either high-power radiatively efficient giant radio galaxies (`high-excitation', HEGRGs) or low-power radiatively inefficient giant radio galaxies (`low-excitation', LEGRGs); the black hole mass (M$_{\rm BH}$); their galaxy star formation rate (SFR); and their molecular gas content. The key to understanding the giant nature of GRGs may lie in the accretion properties of these peculiar AGNs, and we can probe this by tracing molecular gas to determine whether these objects are fueled by cold gas in the disc. 
The presence or absence of cold molecular gas in the host galaxies of GRGs can provide information about their star formation rate, accretion state, and stellar mass properties.

This paper is organised in the following way: In Section.\ \ref{sample} we describe our sample, observations are presented in Section.\ \ref{obs}, and the results are reported in Section.\ \ref{res}. The interpretation of these results is presented and their effect on the evolution scenarios is discussed in Section.\ \ref{disc}. The conclusions of our work are presented in Section.\ \ref{sum}. To compute distances, we have adopted a flat $\Lambda$CDM cosmology, with $\Omega_{\rm m}$=0.29, $\Omega_\Lambda$=0.71, and the Hubble constant of H$_0$ = 70 \kms Mpc$^{-1}$. 

\section{The sample}\label{sample}
\begin{figure}[ht]
  \centerline{
\includegraphics[scale=0.14]{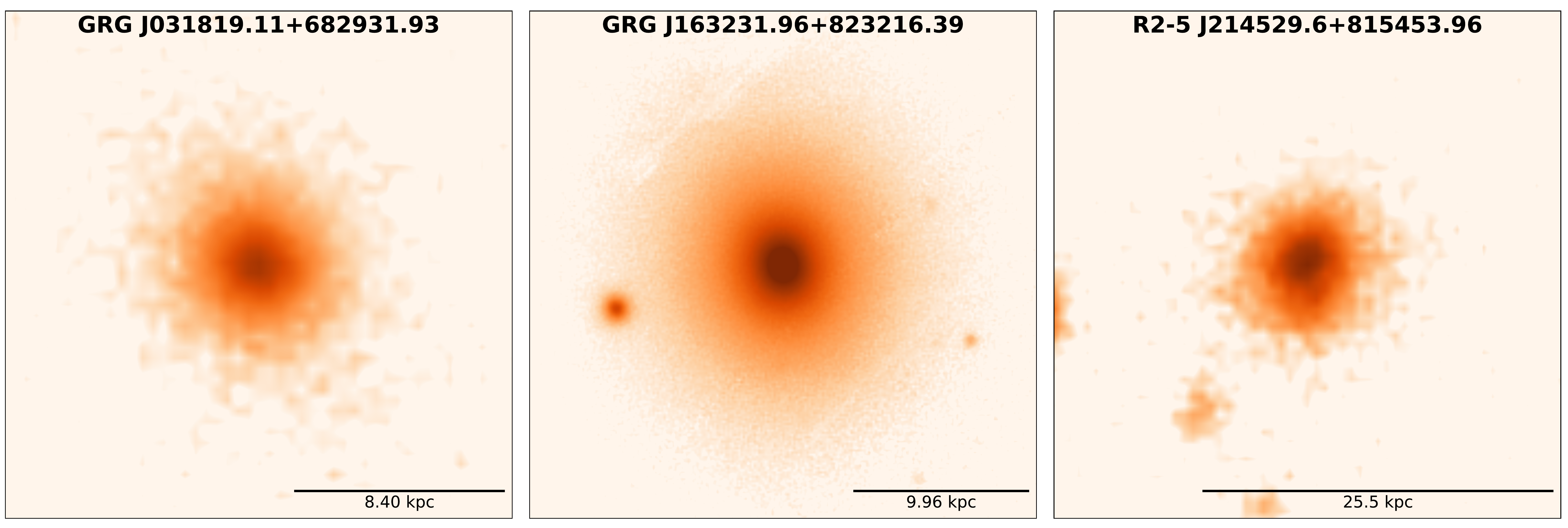}
}
  \caption{PanSTARRS i-band image of the hosts of GRGs that are
    absent in the SDSS footprint. The bar at the bottom indicates the angular scale
  of 5\arcsec, 20\arcsec , and 10\arcsec \ in kiloparsec (kpc) for R2-1, R2-4, and R2-5, respectively.}
\label{fig:optical}
\end{figure}

With the project SAGAN \citep{DabhadeSAGAN20}, we have created the largest database of GRGs to date, and we applied the following criteria to form a sub-sample to be studied at millimetre waves using IRAM-30m \citep{baars87iram} to probe the cold molecular gas content:
(i) the GRG is not hosted by a quasar, (ii) the GRG is well detected 
in the Wide-field Infrared Survey Explorer 
(WISE)  at 22$\rm \mu$m (fourth$^{\rm }$ band) \citep{Wright2010}, (iii) the redshift of the host galaxy of the GRG is accurately known from spectroscopic 
measurements, and (iv) the GRG is observable in the sky of IRAM-30m telescope.
GRGs, like RGs, are mainly early-type galaxies, and only a minority of them ($\sim$17\%, \citealt{Koziel2020}) are late types and may still have some gas; the WISE selection is there to find this minority.
After applying all the above filters to our GRG-catalogue, we selected a sample of 
12 GRGs that is representative of this minority; the GRGs are listed in Table.\ \ref{tab:sample}.
In Cols. 3 and 4 of Table.\ \ref{tab:sample} we 
list the accurate position (RA and Dec) of the host galaxies, and their accurate 
spectroscopic redshifts are listed in column 5. Column 6 includes the luminosity distances of the sources. The information obtained from the WISE fourth$^{\rm }$ band  
and its derived properties are presented in columns 7 and 8, where in column 8, 
L$_{\rm 22 \mu m}$ is the luminosity of the host galaxies at 22$\rm \mu m$. Column 9
contains the GRG references of the sources and/or the optical identification.
Figure.\ \ref{fig:SDSS} displays the SDSS images for the sources when they are available. For the others, the PanSTARRS i band images of the hosts of the 
GRGs in our sample are given in Figure.\ \ref{fig:optical}.

For all the 12 galaxies of the sample, the star formation rate (SFR) was first
estimated from the 22$\mu$m luminosity \citep{Calzetti2007}. When a more accurate
estimate existed in the literature from the SED or multi-wavelength tracers,
we used it from \cite{Chang2015} for R1-2, R1-3, R2-0, R2-3, R2-6 to R2-9, and from 
\cite{Salim2018} for R2-0, R2-3, R2-7 and R2-8. Because several
estimates are available, we took an average value. 
SFR tracers in general suffer from uncertainties, but the SFR estimated from 22$\mu$m was found to be compatible within the uncertainties with the value derived from the FIR when available. Moreover, recent studies of \citet{CLUVER14}, \cite{gurkan15}, and \citet{califa15} have used 22$\mu$m from WISE to determine the SFR effectively. Although the 22$\mu$m flux could also come from the AGN, it is statistically well correlated to the more specific far-infrared SFR indicator, even for AGN \citep{gurkan15}. 
Stellar masses were estimated from observed optical and near-infrared (NIR) magnitudes, using standard relations as a function of colours, derived from stellar population models \citep{Bell2003}. 
Because the targets have a low redshift, K-corrections were applied using
the analytical approximations from \cite{Chilingarian2010}. All adopted
values are listed in Table.\ \ref{tab:char}, together with the CO results we obtained using IRAM-30m data.
Table.\ \ref{tab:char} also lists the angular 
sizes (column 2) of the galaxies, acquired by us from the i-band stellar profiles of the galaxies: ten were from the SDSS, and four from PanSTARRS, see
 Figures.\ \ref{fig:SDSS} and \ref{fig:optical}.

\begin{table*}[h!]
      \caption[]{General properties of our sample of 12 GRGs. Here R1 and R2 refer to observation runs 1 and 2 with IRAM-30m. RA and Dec are expressed in the J2000 coordinate system. Object R1-0 is a radio-loud quasar, and object R2-0 is a radio galaxy. These two objects were recently classified as non-GRGs based on new available data. All the redshifts listed in column (5) are from spectroscopic measurements, where redshifts marked with a dagger$^{\text{}}$  are taken from the SDSS. In the absence of SDSS spectroscopic redshift measurement, the redshift was taken from the reporting paper (Column (9)). Columns (7) and (8) represent the WISE band 4 magnitude and luminosity, respectively. }
         \label{tab:sample}
\centering
\begin{tabular}{ccccccccc}
\hline
Sr.No&Name & RA    &    Dec & $z$ & D$\rm _L$& W4 (22$\mu m$)  &Log L$_{\rm 22 \mu m}$ & Ref \\
    &      & (HMS) &  (DMS)    &  &  (Mpc) &  mag (snr)  & (erg s$^{-1}$) &  \\
  (1) &   (2)   & (3) &  (4)    & (5) &  (6) &  (7)  & (8) &  (9) \\ \hline
\hline
\\
R1-1& 2MASX J23453268$-$0449256 &23 45 32.70 & $-$04 49 25.4 & 0.07557& 342&7.135 (8.5)&43.35 &(1) \\ 
R1-2&CGCG 245$-$031&13 12 17.00 & +44 50 21.3 & 0.03557$^{\dagger}$& 156 &8.133 (5.7)&42.27 &(2) \\ 
R1-3&B2 1029+28 &10 32 14.02 & +27 56 01.7  &0.08519$^{\dagger}$& 389&6.821 (11) &43.58 &(2) \\

R2-1&WNB0313+683&03 18 18.98 & +68 29 31.4 &0.09010& 412 & 6.187 (23.9)&43.18 &(3) \\ 
R2-2&LCRS B132559.3$-$025215  &13 28 34.37 & $-$03 07 44.8 &0.08526$^{\dagger}$& 389 &7.067 (13.5)&42.77 &(4)\\ 
R2-3&2MASX J14504940+1006497  &14 50 49.40 & +10 06 49.1 &0.05453$^{\dagger}$& 243 &8.112 (4.9)&41.95 &(5) \\ 
R2-4&NGC 6251    &16 32 31.97 & +82 32 16.4 &0.02471& 108 &5.561 (33) &42.24 &(6) \\ 
R2-5&LEDA 2785926&21 45 30.90 & +81 54 53.7 &0.14570& 692 &6.320 (28.3)&43.57 &(7) \\
R2-6&B2 0951+27  &09 54 19.19 & +27 15 59.9 &0.47120$^{\dagger}$& 2649 &7.469 (7.0)&44.26 &(8,9) \\
R2-7&LEDA 1406818  &10 21 24.21 & +12 17 05.4 &0.12940$^{\dagger}$& 606 &8.248 (3.5)&42.66 &(10) \\
R2-8& Speca    &14 09 48.85 & $-$03 02 32.5 &0.13749$^{\dagger}$& 651& 8.332 (2.2)&42.69 &(11) \\
R2-9&J121615.21+162432.1&12 16 15.21&+16 24 32.2&0.45908$^{\dagger}$& 2569 &7.556 (6.3)&44.19 &(12) \\
\hline
\\
R1-0&PKS 0211$-$031&02 13 47.00 & $-$02 56 37.5 &0.35679& 1906 &7.412 (8.8)&44.73 &(13)\\ 
R2-0&B2 0915+32B (NE)&09 18 59.41 &+31 51 40.7&0.06212& 278 &8.582 (3.3)&41.85 &(14) \\ 

\hline
\end{tabular}
\begin{list}{}{}
\item -- References 
  (1) \cite{Bagchi2014} --
  (2) \cite{sch01} --
  (3) \cite{Schoenmakers1998}  --
  (4) \cite{machalski01}  --
  (5) \cite{Clarke2017}   --
  (6) \cite{Waggett1977}  --
  (7) \cite{Schoenmakers2000} --
  (8) \cite{Proctor2016} --
  (9) \cite{amir16} --
  (10) \cite{koziel11} --
  (11) \cite{Hota2011} --
  (12) \cite{DabhadeSAGAN20} --
  (13) \cite{Owen1995} --
  (14) \cite{Kuzmicz2018}.

\end{list}
\end{table*}

\subsection{R1-1: 2MASX J23453268$-$0449256}

The galaxy 2MASX J23453268$-$0449256 (R1-1) is one of the largest (projected linear size $\sim$1.6 Mpc at radio wavelengths) radio galaxy known to date, hosted by a massive, late-type spiral galaxy that lacks a classical bulge at the redshift of 0.0755 \citep{Bagchi2014}. 
It shows FR-II and double-double (two sets of lobes, indicating episodic activity) radio morphology; the jet axis is nearly orthogonal to the disc of the host galaxy. This is very unusual because such large-scale radio jets are in general hosted by ellipticals \citep{Taylor1996, McLure2004}. It shows several other unusual properties that challenge the standard paradigms for black hole growth
in galaxies lacking classical bulges. Its general properties as presented by \citet{Bagchi2014} and Bagchi et al. (2020 - in preparation) are similar to those of passive red spirals \citep{Masters2010}, although the  WISE mid-infrared (mid-IR) colours suggest that star formation is still occurring within its disc. With M$_r \sim$ $-$23.0 mag, it lies at the very top of the luminosity
function of red spirals. The galaxy rotation curve derived from the H$\alpha$ line peaks at about 429 $\pm$ 30 km s$^{-1}$ at a large radius, which is quite
high for spirals, after correcting for i = 65$^{\circ}$  inclination angle (cf. HyperLeda\footnote{\url{http://leda.univ-lyon1.fr/}}). This implies an estimated dynamical mass of $\sim$10$^{13}$ M$\rm _\odot$ within
the virial radius of 450 kpc, assuming a flat rotation curve. The central bulge velocity dispersion is $\sigma$ = 351 $\pm$ 25 km s$^{-1}$, which is close to the highest velocity dispersion found among nearby E and S0 galaxies \citep{vandenBosch2012}. \citet{Walker2015}
used Chandra to detect the hot X-ray halo of the galaxy, which extends 
to r $>$ 80 kpc; it is one of just a few high-mass spirals for which hot gaseous haloes have directly been detected through X-ray imaging to date \citep{bregman18}.
Intriguingly, the X-ray emitting hot halo is highly elongated parallel to the host galaxy disc, suggesting that feedback by the radio jet has disrupted or expelled hot gas from a region extending roughly 80 kpc
above and below the disc plane. Radio-mode feedback like this is rarely seen in spiral galaxies \citep{dipanjan16,dipanjan18}.

\subsection{R1-2: CGCG 245$-$031}
CGCG 245$-$031 (R1-2) is a flat-spectrum radio source \citep{Healey2007}. It has a low value for its integrated spectral index and is a young radio source. It is a good candidate for a restarting class.

\subsection{R1-3: B2 1029+28}
B2 1029+28  is hosted by a Seyfert type-1 galaxy \citep{veron06} called 2MASX J10321402+2756015. In this galaxy, the stellar populations are not old, but have intermediate ages \citep{Kuzmicz2019}.

\subsection{R2-1: WNB0313+683}
The GRG R2-1 or WNB0313+683 has a projected linear size of 1.52 Mpc with a GHz-peaked spectrum (GPS) core, FR-II type morphology, and recurrent activity \citep{Schoenmakers1998, Bruni2019}. It is hosted by a Seyfert type-1 galaxy (LEDA 3095635) at $z \sim$0.09 that is detected in hard X-rays \citep{winter08}. \cite{Schoenmakers1998} estimated its spectral age to be $\sim$ 144 Myr, but recently, \citet{Machalski11} estimated the age of the lobes to be $\sim$250 Myr.

\subsection{R2-2: LCRS B132559.3$-$025215}
\citet{m99} identified LCRS B132559.3$-$025215 as a GRG. It is hosted by the Seyfert 2-type \citep{veron06} galaxy called 2MASX J13283434$-$0307442.

\subsection{R2-3: 2MASX J14504940+1006497}
Source R2-3 is hosted by a central dominant galaxy (the core of a triplet, which is also known as UGC 09555 NED03) of a low-density galaxy group (MSPM 02158), as shown in Figure.\ \ref{fig:SDSS}. Based on their SED analysis, \citet{Clarke2017} concluded that the host galaxy is an S0-type galaxy with an SFR of 1.2 $\pm$ 0.3 M$\odot$ yr$^{-1}$. \citet{Clarke2017} also presented its LOFAR 142 MHz radio map, which reveals a large diffuse structure that is not seen in any other radio survey. The bright radio core and high diffuse emission along the jet axis strongly indicate a restarted source.
Based on its radio core spectrum, it has been classified as a flat-spectrum galaxy \citep{Healey2007} with frequencies between $\sim$1 to 10 GHz, and it has an inverted core spectrum with frequencies between 22 to 43 GHz \citep{parkkvlbi13}. This further supports the hypothesis that this is a restarted AGN.

\subsection{R2-4: NGC 6251} 
 NGC 6251 is our nearest target, and it has been the object of many
observations since its identification \citep{Waggett1977,Laing1983}.
\cite{Chen2011} studied the neighbouring 17 galaxies, which may form a loose
group, but with very low background X-ray gas, 
explaining the large 1.6 Mpc radio jets.
The main jet and lobes are centre-brightened like FR-I,
but there are also hot spots near the end of the north and south lobes
that suggest a FR-II type of morphology. The radio power is in between the two FR types.
Recent LOFAR observations at 150 MHz have revealed low surface brightness extensions
and backflows of the lobes \citep{Cantwell2020}, with ages older than 200 Myr.
A dust lane has been discovered by \cite{Nieto1983}, roughly parallel
to the radio jet, with a size of about 5 kpc. The inner orientation of the jets
is still unclear; it appears at least that
the inner jet and outer lobes are misaligned \citep{Jones2002}.

\subsection{R2-8: Speca}
Only about$\text{ ten}$ RGs \citep{ledlow01,Hota2011,Bagchi2014,Singh2015,mao16,mulcahy16} have been found to be hosted by spiral galaxies up to now, and only two (2MASX J23453268$-$0449256 and Speca) of them belong to the GRG subclass. Along with its unusual property of being hosted by a spiral \citep{Hota2011}, Speca is classified as the central galaxy, called MaxBCG J212.45357-03.04237 cluster or group. Another peculiar property of this object is its three pairs of lobes, also called a triple-double, with clear evidence of restarting activity. The outer lobes look like relics revived by shocks coming from the interaction between the cluster and cosmic filaments. The galaxy shows very recent star formation that might be fueled by a cooling flow from merging activity $\sim$500 Myr ago \citep{Hota2011}.

\subsection{Other GRGs}
R2-5 (Seyfert 2) and R2-7 are also dominant galaxies of their groups, with indications of star formation in the recent past \citep{Kuzmicz2019}. R2-9 is one of the most recently identified GRG, at relatively high redshift \citep{DabhadeSAGAN20}. It appears to be actively forming stars. R2-9 is the most luminous GRGs at 22 microns (WISE band 4) in our sample, and appears to be harbouring a dust-obscured powerful AGN based on its red colour.

\subsection{Non-GRGs}
At the beginning of our study, two objects were thought to be GRG and were observed at IRAM. We report the results, which may serve other studies, although the objects are not GRGs, but normal RGs or RQs.
B2 0915+32B is a galaxy pair, and the two companions lie at a distance of 43~kpc. The early-type northern galaxy is the host of the giant radio jet, and
its southern companion is a late-type blue spiral galaxy, which could be the reservoir of gas to fuel the northern AGN. PKS~0211-031 is radio-loud quasar with FR-II type morphology.

\begin{figure}[ht]
\centerline{
\includegraphics[scale=0.345]{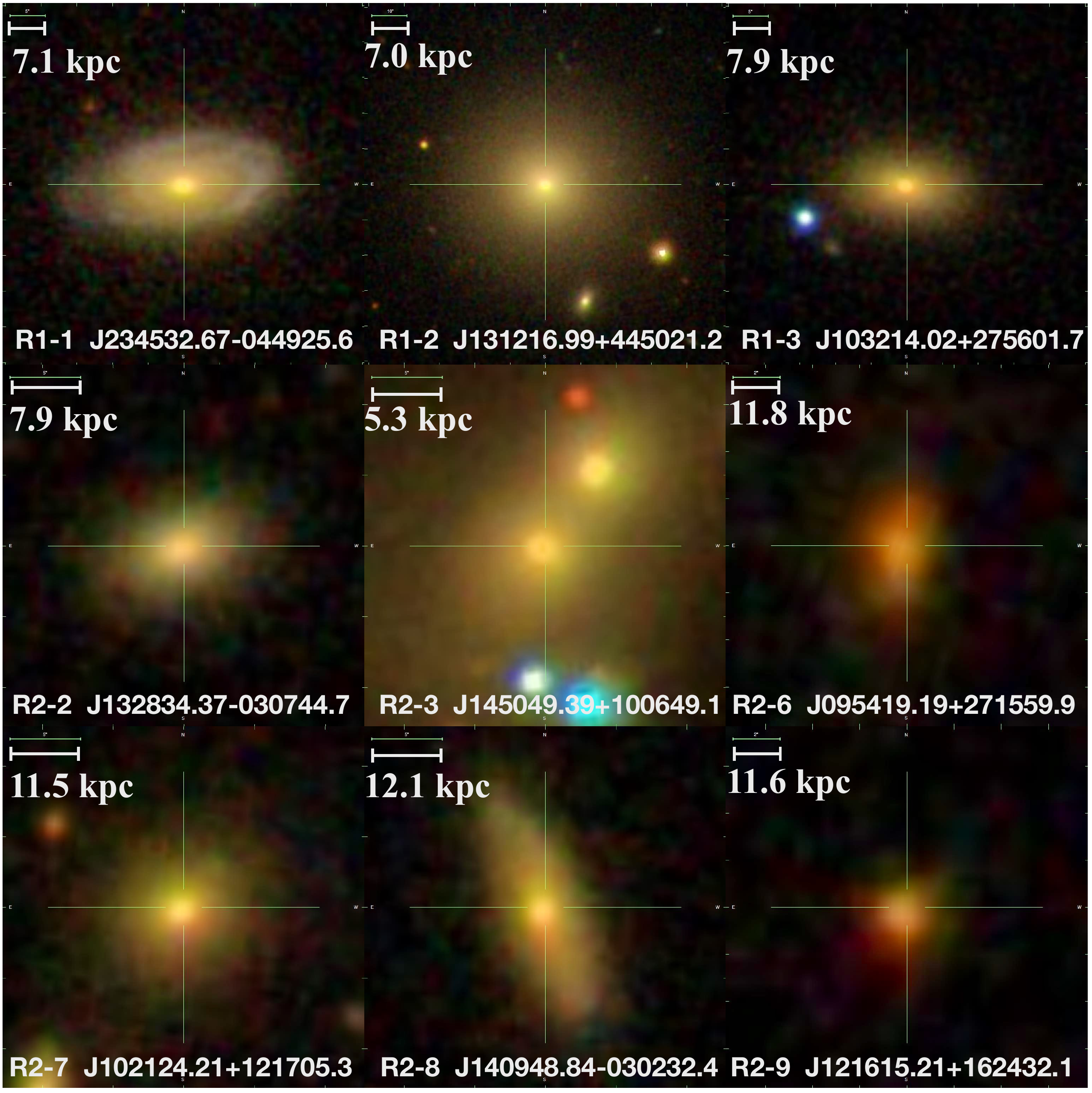}
}
\caption{Host galaxies of nine GRGs from SDSS. The bar at the top left represents the angular scale in kpc for reference.}
\label{fig:SDSS}
\end{figure}

\section{Observations} \label{obs}
Observations of the CO(1-0), CO(2-1) (and for one target CO(3-2)) emission lines have been carried out with the IRAM-30m telescope at Pico Veleta, Granada, Spain, during June and September 2016 for 
the first run (called R1), and then in March 2020 for the second run (called R2). The beam full width at half-maximum (FWHM) is 23\arcsec and 12\arcsec at the frequencies of 115 GHz and 230 GHz, respectively. The redshifts of our targets (except for two objects) range from 0 to 0.1, and the corresponding beams are up to 10\% larger, accordingly.
The SIS receivers (EMIR) were used for observations in the wobbler-switching mode, and the reference positions were offset by $\pm$120 arcsec in azimuth. The main-beam efficiency of IRAM is $\rm \eta_{mb}=T_{A}^{*} / T_{mb}$ = 0.79 and 0.63 at 105 GHz and 210 GHz, respectively. The system temperatures ranged between 150 K and 400 K at 2.6 mm, and between 200 K and 600 K at 1.3 mm. The pointing was regularly checked every two hours on a nearby planet or a bright continuum source, and the focus was reviewed after each sunrise or when a suitable planet was available, as well as at the beginning of each day. The time on source typically ranged from four to five hours, depending on the weather.
Two backends were used simultaneously, the Wideband Line Multiple Autocorrelator (WILMA) and the the Fourier transform spectrometer (FTS). The $\rm T_{\rm mb}$ rms noise level in four hours of integration was $\sigma _{\rm mb}\sim$0.7mK with a spectrometer resolution of 50 km s$^{-1}$ for 105 GHz and $\sigma_{\rm mb}\sim$0.9mK for 210 GHz. These can be lower in very good weather.
The upper limits reported in the next section for the integrated CO fluxes are computed  by S(CO)dV = 3 $\times$ rms$_{300}$ JypK $\times$ 300 km s$^{-1}$ , where JypK is the conversion from K to Jy (=5 Jy K$^{-1}$ for the IRAM-30m), and rm$_{300}$ is the rms estimated in 300 km s$^{-1}$ wide bins. 
A typical FWHM of 300 km s$^{-1}$ is usually taken for early-type galaxies, which RGs and GRGs are, as reported in the catalogue by \cite{Young2011}. 

Two of our sample sources have previously been observed with the IRAM-30m by \citet{TANGO2010}. For NGC 6251, we combined our spectra in CO(1-0) to improve the upper limit and provide a CO(2-1) search, which was lacking. As for B2 0951+32B, which is a pair of galaxies with a projected separation of 36\arcsec, the previous pointing was  directed towards the NED middle position in between the two galaxies, and did not cover either.  We observed towards the north-eastern elliptical galaxy, which is the host of the RG \citep{Fanti1987}.

\begin{figure}
\centering
\includegraphics[scale=0.58]{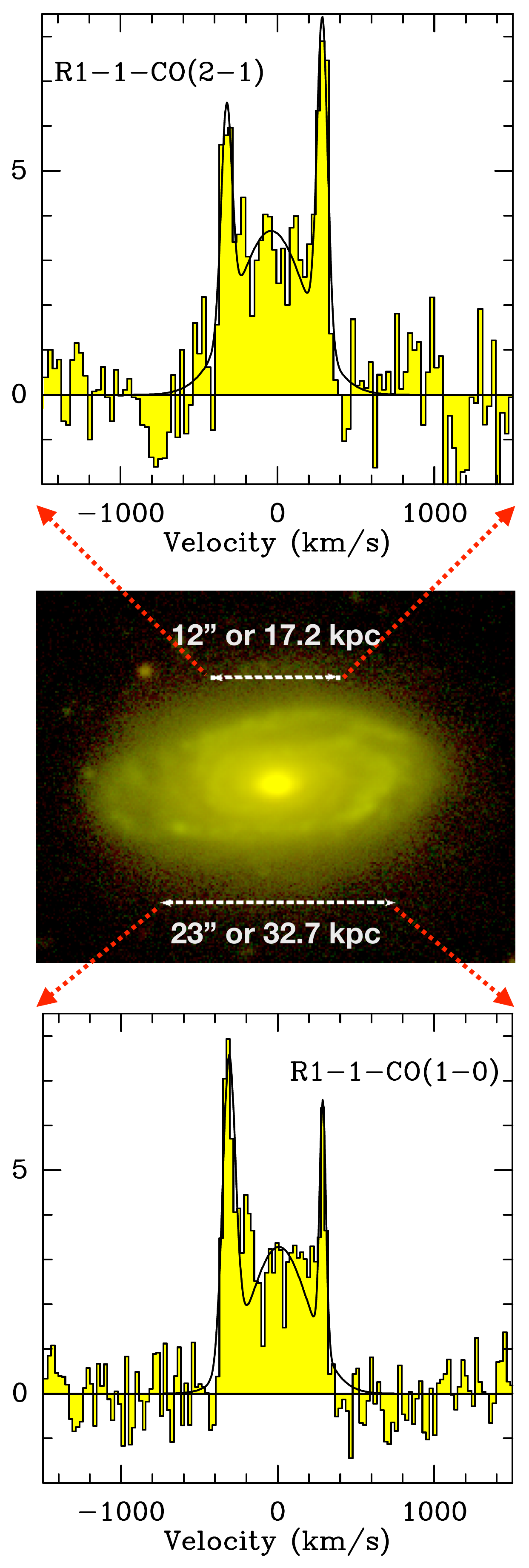} 
\caption{Two spectra obtained towards R1-1 or GRG-J2345$-$0449 in CO(1-0) (bottom) and CO(2-1) (top). The vertical scale is T$\rm_{mb}$ in mK. The middle optical image of the host spiral galaxy of J2345$-$0449 is obtained by combining G- and R-band data of MegaCam on the CFHT. The G- and R-band data belong to the first-generation filters on the MegaCam. The scale on the image represents the beam of the IRAM observations at the respective frequencies.}
\label{fig:IRAM}
\end{figure}

\begin{figure*}
\centering
\includegraphics[scale=0.92]{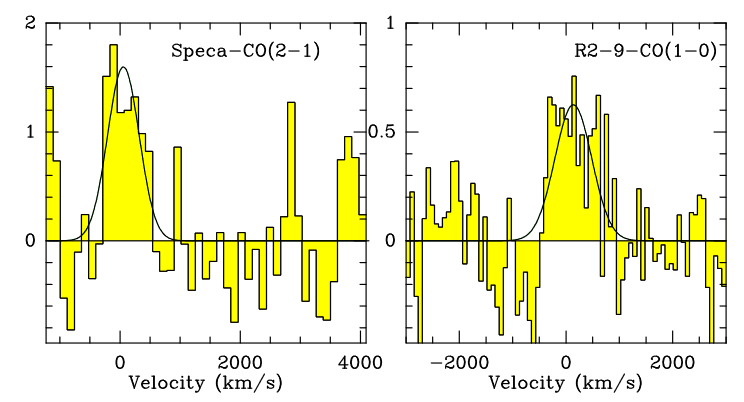} 
\caption{Tentative spectra obtained towards Speca in CO(2-1) (left) and R2-9 in CO(1-0) (right). The vertical scale is T$\rm _{mb}$ in mK. The CO(2-1)/CO(1-0) line ratios for the two objects are within reasonable expectations, as shown in Table \ref{tab:char}.}
\label{fig:tentative}
\end{figure*}

\section{Results}
\label{res}

As shown in Table.\ \ref{tab:char}, only the spiral galaxy R1-1 (J2345$-$0449) has been clearly detected, and upper limits are reported for the 11 other GRGs, with tentative detections for 2 of them.
The two-horn spectra of R1-1 (Figure.\ \ref{fig:IRAM}) were fitted by
three Gaussians each, and the fits are reported in Table.\ \ref{tab:fit}. These fits are only empirical, which is better for computing the total intensity, but do not imply any underlying physical components.
The upper limit for the R1-0 RG (PKS0211$-$031) is relatively high because it is a source with one of the highest redshifts. However, the molecular mass is expected to increase with redshift \citep{Tacconi2018}, and based on the measured SFR, a much higher molecular mass was expected. Therefore the upper limit is significant. The tentative signals are displayed in Figure \ref{fig:tentative} for CO(2-1) in the spiral galaxy Speca, and for CO(1-0) for J121615.21+162432.1. Only linear baselines were subtracted from the spectra, but uncertainties are still present because of the presence of wavy baselines, although the signal appears at 5$\sigma$ and 6$\sigma$ , respectively.

In the following subsections, we present the results we derived from our detections and upper limits of the molecular gas content of the hosts of GRGs, and also compare it with the SFR derived from the mid-IR fluxes.
All our upper limits are significant, as we discuss in Sec.\ \ref{disc} and show in particular in Fig. \ref{fig:SFR-KS} and Fig. \ref{fig:fgas-tdep}. In these scaling relations, the upper limits are within or below the expectations.
Using our observed quantities, we were also able to estimate the gas depletion time t$\rm _{dep}$ = M(H$_2$)/SFR and compare it with the radiative lifetime of the GRGs. From the shape of the detected profile (two-horn shape, and contrast), we derived the gas concentration in the host and discuss the origin of the cool gas. The two-horn profile indicates an extended disc or ring, which is relaxed in the potential, suggesting a quite old galaxy accretion or merger origin for the gas. When the gas comes from cooling filaments alone, it generally has no time to relax in the galaxy potential, and no such profile is expected.

\begin{table*}[h!]
      \caption[]{Derived properties of the GRG sample, where R$_e$ is the half-light radius, SFR is the star formation rate, CO refers to  CO flux in the respective bands, log M(H$_2$) represents the molecular mass (log scale), t$_{\rm dep}$ is the depletion time, and logM$_*$ refers to the stellar mass (log scale) of the galaxy.}
         \label{tab:char}
\centering
\begin{tabular}{ccccccccc}
\hline
Sr.No & R$_e$ &  R$_e$ &SFR  & CO(1-0) & CO(2-1) & log M(H$_2$) & t$_{\rm dep}$ & logM$_*$\\
&(arcsec)&(kpc)&($\rm M_{\odot} \ yr^{-1}$)&(Jy\kms)&(Jy\kms )&(M$\rm _\odot$)& Gyr&(M$_\odot$)\\ 
 (1) &   (2)   & (3) &  (4)    & (5) &  (6) &  (7)  & (8) &  (9) \\
\hline
\hline
\\
R1-1        &5.4 &7.76& 2.95&   14.0   &  16.3  & 10.21  &5.49   &11.37 \\ 
R1-2        &6.1 &4.32& 0.19&   $<$1.7 &  $<$2.7&$<$8.63 &$<$2.24&10.51 \\ 
R1-3        &3.6 &5.77& 2.45&   $<$1.5 &  $<$7.5&$<$9.35 &$<$0.91&10.84 \\ 
R2-1        &0.96&1.62& 2.07&   $<$1.3 &  $<$2.2& $<$9.34&$<$1.05&10.63 \\ 
R2-2        &2.2 &3.53& 0.91&   $<$0.9 &  $<$1.3& $<$9.13&$<$1.48&10.52 \\ 
R2-3        &0.78&0.83& 0.09&   $<$0.9 &  $<$1.9& $<$8.73&$<$5.96&11.25 \\ 
R2-4$^{(2)}$&15.6&7.78& 0.31&   $<$1.2 &  $<$4  & $<$8.16&$<$0.47&11.40 \\ 
R2-5        & 0.7&1.79& 4.64&   $<$1.1 &  $<$1.9& $<$9.69&$<$1.06&11.03 \\ 
R2-6$^{(1)}$&0.95&5.63&  170&   $<$1.3 &  $<$2.7&$<$10.82&$<$0.39&11.70 \\ 
R2-7        & 2.4&5.59& 1.44&   $<$1.1 &  $<$1.9& $<$9.59&$<$2.70&11.41 \\ 
R2-8        & 3.3&8.04& 1.48&   $<$0.9 &     1.1& 9.04   &  0.74&11.50 \\ 
R2-9$^{(1)}$& 0.6&3.55& 8.33&   0.54  & $<$1.1 & 10.42  & 3.15 &10.89 \\ 
\hline
R1-0$^{(1)}$&0.87&4.37&48.9  &  $<$0.45& $<$1.6 &$<$10.11&$<$0.26&10.87 \\ 
R2-0        &5.6 &6.70& 0.47&   $<$1.1 &  $<$2.7& $<$8.93&$<$1.81&11.46 \\ 

\hline
\end{tabular}
\begin{list}{}{}
\item  (1) For these sources, we observed CO(3-2) instead of CO(2-1).
(2) This source was also observed by \cite{TANGO2010}.
 All upper limits are computed at 3$\sigma$ and assuming an FWHM = 300 \kms.
\end{list}
\end{table*}

\begin{table}[h!]
      \caption[]{Fit results for R1-1 (GRG-J2345$-$0449) CO spectra.}
         \label{tab:fit}
\centering
\begin{tabular}{ccccc}
\hline
        Line & Area$^{(1)}$ & V  & FWHM & T$_{\rm mb}$ peak \\
           &  mK \kms &  \kms & \kms& mK\\ \hline
\hline
CO(1-0) &1232$\pm$167&-316$\pm$4  & 207$\pm$11 & 5.6$\pm$0.8 \\
CO(1-0) &800$\pm$121 &  10$\pm$18 & 235$\pm$50 & 3.2$\pm$0.8 \\
CO(1-0) &779$\pm$150 & 291$\pm$15 & 179$\pm$30 & 4.1$\pm$0.8 \\
\hline
CO(2-1) &930$\pm$115 &-324$\pm$8  & 141$\pm$20 & 6.2$\pm$0.9 \\
CO(2-1) &1368$\pm$265& -35$\pm$33 & 357$\pm$70 & 3.6$\pm$0.9 \\
CO(2-1) &972$\pm$110 & 286$\pm$5  & 113$\pm$13 & 8.1$\pm$0.9 \\
\hline
\end{tabular}
\begin{list}{}{}
\item  (1) Integrated signal in the main-beam temperature scale.
\end{list}
\end{table}

\subsection{Molecular masses}
\label{masses}

To quantify the amount of molecular gas found in each target, we relied on the empirically established proportionality between the CO(1-0) luminosity and H$_2$ mass for a large number of Milky Way-like galaxies \citep[e.g.][]{Bolatto2013}. We converted the integrated intensities in each beam (in T$_{\rm mb}$) into integrated fluxes S(CO)dV, with a conversion of 5 Jy per Kelvin, as appropriate for the IRAM-30m telescope.
From the integrated CO(1-0) flux S(CO)dV (Jy \kms) found within each region, the molecular mass was obtained through the formula \citep[e.g.][]{Solomon2005}
\begin{equation}
    {\rm L'}_{\rm CO} (\rm K \kms \rm pc^{-2}) = 3.25 \times 10^7 \frac{\rm S(CO) dV}{(1+z)}  \left(\frac{\rm D_L}{\nu_{\rm rest}}\right)^2
,\end{equation}

where $\nu_{\rm rest}$ =115.271 GHz, and D$\rm _L$ is the luminosity distance in megaparsec. The molecular mass, including helium, was then derived from ${\rm M}(\rm H_2)$ = 4.36 ${\rm L'}_{\rm CO}$ assuming the standard CO-to-H$_2$ conversion factor of X$_{\rm CO}$ = 2 $\times$ 10$^{20}$ cm$^{-2}$(K~\kms)$^{-1}$, as applicable to Milky Way-like galaxies. The region encompassed by the CO(1-0) beams in each galaxy is sufficiently large to cover the entire CO emission (even in the nearest galaxy, NGC~6251 (R2-4), it is 11 kpc). It is, therefore, possible to compare the total molecular masses with other published samples without any aperture correction.

\subsection{Deconvolving the two-horn spectra}
\label{model}

\begin{figure*}
\centering
\includegraphics[scale=0.85]{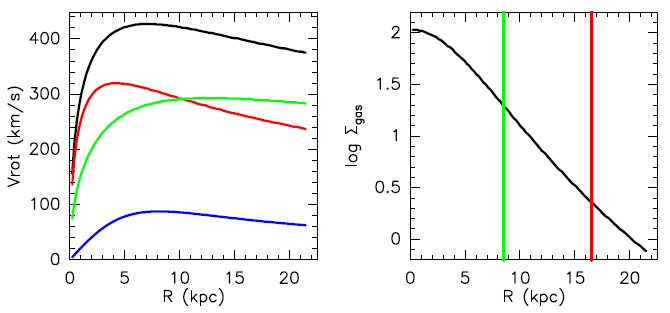} 
\caption{Mass model for R1-1 or GRG-J2345$-$0449 (left), with the contributions to the rotational velocity of the stellar mass (red), the dark matter (green), and the gas (blue). The right panel shows the adopted gas surface density. The two vertical lines indicate the FWHM beam of the CO(1-0) (red) and CO(2-1) (green).}
\label{fig:vrot}
\end{figure*}

In Figure.\ \ref{fig:IRAM} the CO(1-0) and CO(2-1) spectra show a two-horn shape, with a different depth between the two peaks; the CO(1-0) spectrum is deeper. This means that the molecular component is not too highly concentrated towards the centre and that we are beginning to resolve it in the CO(2-1) line. Because the rotational velocity is zero at the nucleus, the central part of the profile is boosted by a high gas concentration. With a large beam, that is, for CO(1-0), the outer parts of the disc are preponderant, and the two-horn profile is more contrasted. The smaller CO(2-1) beam begins to resolve the disc, and the central part is more weighted than the outer parts in this line profile. 
To interpret the kinematics of our CO data more quantitatively, we used a simple analytical model that computes
the velocity spectrum from the rotation velocity profile of the galaxy \citep{Wiklind-1997}. The rotation
velocity profile is determined from the stellar mass distribution,  assuming it follows the simple potential-density pair proposed by \citet{Hernquist1990}. We adjusted the size and mass of the stellar component to fit the half-light radius  and M$_*$ as shown in Table.\ \ref{tab:char}. We completed this mass at high radii by dark matter, in order to flatten the rotation curve. The dark matter mass inside 24.5 kpc was taken to be 1.5 M$_*$, and its characteristic radius was three times that of the stars. The gas disc was also taken into account in the potential, as shown in Figure.\ \ref{fig:vrot}, with the mass shown in Table.\ \ref{tab:char}.
Its characteristic scale can be determined by modelling the two-horn spectra. Based on the ellipticity obtained from the optical image (Figure.\ \ref{fig:SDSS}), we derive an inclination of 65$^\circ$, which is also close to the value given by Hyperleda (65.7$^\circ$).

The gas distribution is assumed to be axisymmetric with a surface density $\Sigma(r)$ 
in a disc with negligible
thickness. For each velocity dv, the code calculates the density contained in 
the isovelocity
(equation \ref{eq:dndv}). The velocity spectrum corresponds to the velocity histogram. We then convolved the computed spectrum with a
Gaussian with a standard $\rm \sigma=10\: km \ s^{-1}$ to account for the gas velocity dispersion,
\begin{equation} \label{eq:dndv}
        \frac{\rm dN}{\mathrm{\rm dv}}(\mathrm{v})=\int \frac{\Sigma(\rm r)\, rdr}{\mathrm{v}_{\rm rot}(\rm r)\sin i \left( 1-[\frac{\mathrm{v}}{\mathrm{v}_{\rm rot}(r)\sin i}]^2 \right)^{1/2}}
.\end{equation}

To determine the concentration of gas in the galaxy, we adopted an analytical profile,
with a potential-density pair, which facilitates computing the rotation
velocity.  We selected the formulation of a Toomre disc of
order 2: $\Sigma(r)=\Sigma_0\left(1+\frac{r^2}{d^2}\right)^{-5/2}$ \citep{Toomre-1964}. 
Figure.\ \ref{fig:vrot} shows that the profile is close to an exponential,
which is realistic. We also tried models with rings, using the difference between
two Toomre discs with different scales, but a single gas disc was an even better fit.
The exponential scale of the best-fit disc is 4.34 kpc, leading to a half-light 
radius of 7.3 kpc, which is very close to that of the stellar component. This corresponds
to what is generally found for the molecular distributions \citep{Lisenfeld2011}.

\begin{figure}[ht]
  \centerline{
\includegraphics[scale=0.54]{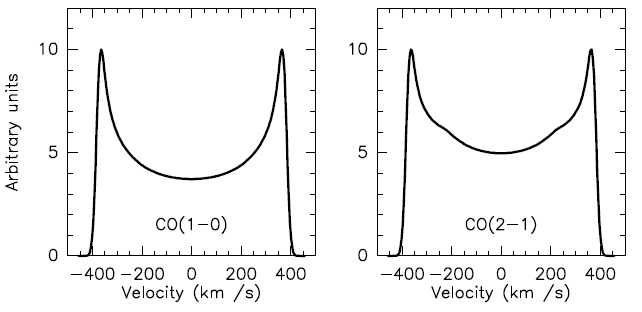}
}
\caption{Spectra obtained from the model for the CO(1-0) line with
a beam of 23\arcsec = 33 kpc (left) and for the CO(2-1) line with a beam of 12\arcsec =17 kpc (right).
The assumed inclination of the disc is 65$^\circ$.}
\label{fig:proj}
\end{figure}

\begin{figure}[ht]
\centerline{
\includegraphics[scale=0.45]{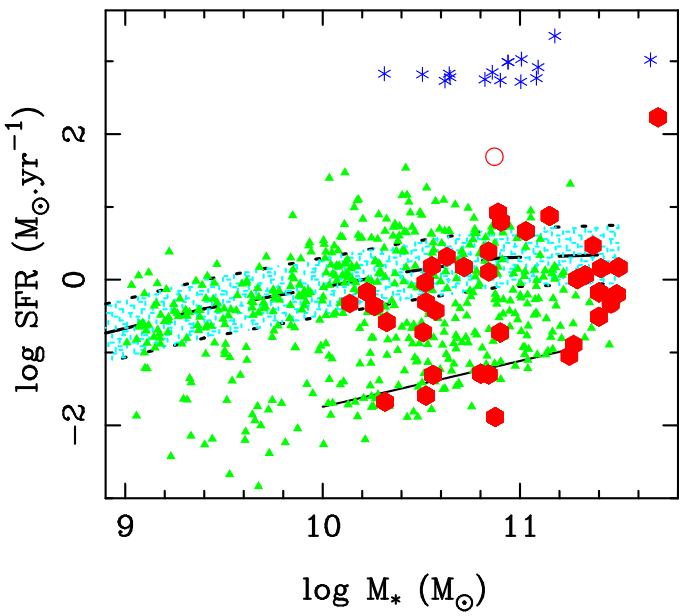}
}
\caption{ Location of our targets (red hexagons) in the SFR-M$_{\rm star}$ diagram.
They are compared with the xCOLD GASS sample of 532 galaxies with local
redshift 0.01 $<$ z $<$ 0.05 from \cite{Saintonge2017} (green triangles), 
        and with the starburst galaxies (blue stars) in the comparable range 
0.2 $<$ z $<$ 0.6 \citep{Combes2011}. The SDSS main sequence is shown with the cyan-shaded 
region, and delimited by dashed
        lines (at $\pm$0.4 dex), and the red sequence by a full line. Open red circles belong to non-GRGs observed by us.
}
\label{fig:MS}
\end{figure}

\begin{table*}[h!]
      \caption{Column (2) lists the half-light radius (R$_e$), column (3) the star formation rate (SFR), column (4) the molecular mass (log M(H$_2$)), column (5) the stellar mass (logM$_*$), column (6) the 22$\mu$m luminosity derived from WISE (log L$_{\rm 22 \mu m}$), and column (7) the CO(1-0) line width (FWHM) obtained from the respective references listed in column (8).}
         \label{tab:other}
\centering
\begin{tabular}{cccccccc}
\hline
Name & R$_e$  & SFR&  log M(H$_2$) & log M$_*$ & log L$_{\rm 22 \mu m}$ & FWHM & Ref \\
& (kpc) &  (M$\rm _\odot$yr$^{-1}$) & (M$\rm _\odot$) &  (M$\rm _\odot$) & (erg s$^{-1}$)&(\kms) &\\ 
(1)&(2)&(3)&(4)&(5)&(6)&(7)& (8) \\
\hline
\hline
NGC315  & 6.3 & 0.63   & 7.87 & 11.49&42.18&150&(1) \\
3C31  & 4.6 & 1.02   & 9.22   & 11.29&41.51&550&(1) \\
3C236 & 7.4 & 7.51   & 9.32   & 11.15&43.07&-&(2) \\
3C326N& 6.9 & 0.67   & 9.18   & 11.40&42.09&351&(3) \\
\hline
\end{tabular}
\tablebib{(1) \cite{TANGO2010}; (2) \cite{Labiano2013}; (3) \cite{Nesvadba2010}}

\end{table*}

%%%%%%%%%%%%%%%%%%%%%%%%%%%%%%%%%%%%%%%%%%%%%%%%%%%%%%%%%%%%%%%

\section{Discussion}
\label{disc}

GRGs are rarer than the RG population, and very few studies have been conducted to detect molecular gas in their host galaxies. We list the detections
reported in the literature in Table.\ \ref{tab:other}. In the following section, we consider the four GRGs listed in Table.\ \ref{tab:other} that are hosted by early-type galaxies. 
NGC~315 is the brightest elliptical (cD) galaxy of a group \citep{Sullivan2015}, and has also been detected in HI-21cm absorption and emission \citep{Morganti2009}. 
3C~31, hosted by NGC~383, is a lenticular galaxy which is well detected in CO \citep{Evans2005, TANGO2010}, but neither in emission nor in absorption \citep{Emonts2010} of HI-21cm.
The host of 3C~236 is a massive elliptical galaxy with a distorted morphology, and a prominent
dust lane. Its CO spectrum is blue-shifted by $\sim$300 km s$^{-1}$ with respect to the 
optical velocity \citep{Labiano2013}, and it corresponds to the peak of the HI-21cm
absorption \citep{Morganti2005}. The oldest and youngest spectral ages of different parts of GRG 3C~236 are 159 Myr and 51 Myr \citep{Shulevski19}.
The radio core of 3C~326 comprises two early-type galaxies, and the most massive galaxy in the north (3C~326N : 15:52:09.10 +20:05:48.32)) is the host galaxy \citep{Rawlings1990}. Its molecular gas in the cold \citep{Nesvadba2010} and warm phase \citep{Ogle2007} traces shocks, possibly due to the radio-jet feedback.

\subsection{Star formation and gas mass}
\label{SFR-gas}

The host galaxies of GRGs have not previously been studied with respect to their SFR as a function of stellar mass and have not been compared with the main-sequence galaxies, starbursts, or quiescent
objects. We show our comparisons in Figure.\ \ref{fig:MS}  , where the xCOLD GASS sample of \citet{Saintonge2017} is a reference for tracing the main sequence at low redshift and as part
of the red sequence. Starburst galaxies at comparable redshifts are plotted from
\citet{Combes2011}. It is important to have samples at the same $z$ 
because the location of the main sequence strongly depends on redshift
\citep{Whitaker2012}.
Eight of our targets are on the main sequence (MS), including the detected R1-1, four are in a starburst phase,
and seven are in the red passive sequence. The position close to the MS
is associated with disc morphology \citep{Wuyts2011}, and the slope and scatter of the MS depend on morphology \citep{Whitaker2015}.

\begin{figure*}[ht]
\centerline{
\includegraphics[scale=0.63]{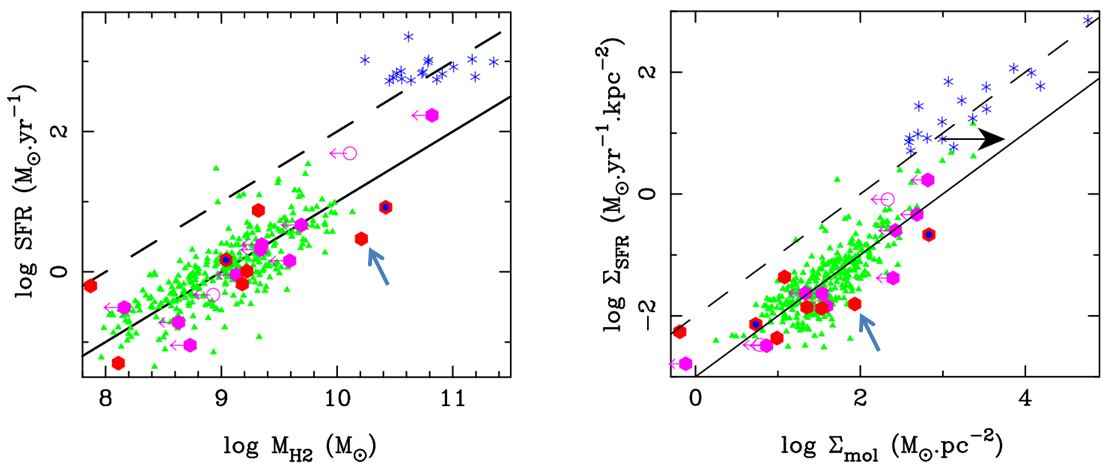}
}
\caption{ Star formation rate as a function of the molecular content
for our targets compared to normal galaxies and starbursts at comparable
redshifts. The global SFR-M$_{\rm mol}$ diagram is shown at the left, and the
surface densities of the same quantities, representing the unresolved Kennicutt-Schmidt relation \citep{Kennicutt1998}, are shown at the right. The straight lines are power laws of slope 1. The full line corresponds to a depletion time of 1~Gyr, and the dashed line to 100~Myr.
Symbols are the same as in Figure.\ \ref{fig:MS}; upper limits of our targets are shown in pink, and
tentative detections as a blue hexagon inside the red one. The blue arrow points to the location of R1-1 (GRG-J2345$-$0449) in the respective plots.}
\label{fig:SFR-KS}
\end{figure*}

The star formation rate is expected to be related to the interstellar gas content, and the precise relationship has been discussed either from the atomic or molecular gas or both \citep{Kennicutt1998, Bigiel2008}. The relation is much tighter and linear with the molecular gas. The relation can be expressed with the global quantities (SFR versus M$\rm _{mol}$), or with the surface densities, when normalised with the effective area. The latter are computed from the effective radius, containing half of the light (displayed in Table.\ \ref{tab:char}). This is a global quantity for the galaxy because we do not resolve the inner parts of the galaxies with our observations with IRAM-30m.
However, the relation between the two average surface densities yields another very useful point of view, as discussed by \cite{Kennicutt1998}.
Figure.\ \ref{fig:SFR-KS} displays both relations. In the global quantity diagram, most of our upper limits fall in the region of normal MS galaxies, traced by the straight line
corresponding to a depletion time t$\rm _{dep}$ = M(H$_2$)/SFR = 1 Gyr. The R1-1 detected
spiral galaxy has a long  t$\rm _{dep}$ of 5.5 Gyr, and the tentative 
detection R2-9 has t$_{\rm dep}$ of 3.15 Gyr, and the two detected GRG (NGC~315 and 3C~236) have an exceptionally short t$_{\rm dep}$ of 120 and 280 Myr, as well as two upper limits that are more
typical of starbursts.  
When normalised to the galaxy surface, the R1-1 spiral galaxy and other outliers fall much closer to the main sequence in the KS diagram, except for NGC~315 and NGC~6251, which are still exceptionally deficient in molecular gas. R1-1, which is a giant spiral galaxy, does not have a high SFR, as seen in Figure.\ \ref{fig:SFR-KS}, although its reservoir of M$\rm
_{H2}$ is large, which may be connected with the restarted nature of the GRG.
In these relations, the molecular
masses of the starburst sample were obtained with a different CO-to-H$_2$ conversion
factor, as is justified for starbursts \citep{Solomon2005}. With a larger 
factor, the difference to the MS would be much smaller, as is indicated
by the arrow in the right panel of Figure.\ \ref{fig:SFR-KS}.

\begin{figure*}[]
\centerline{
\includegraphics[scale=0.61]{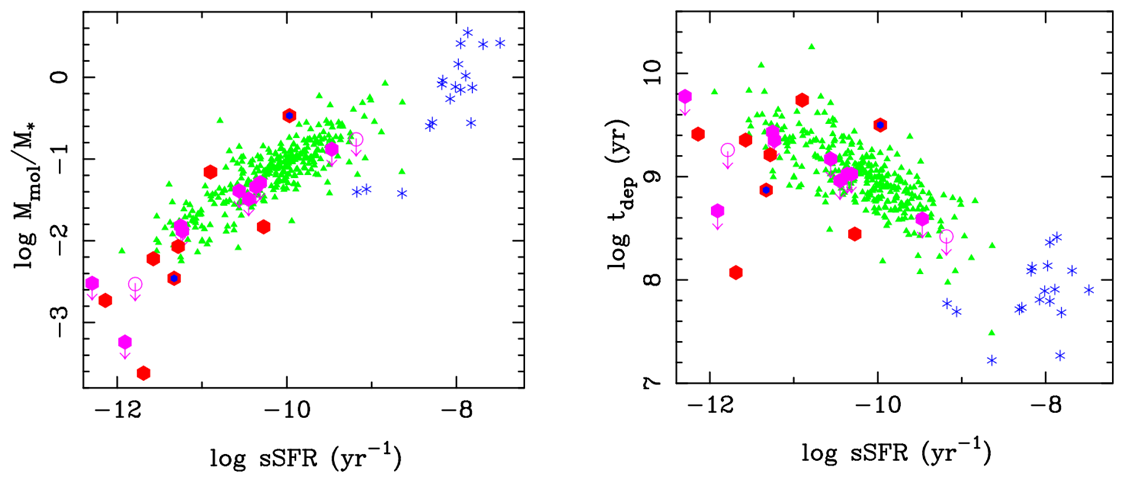}
}
\caption{Trends of specific star formation with other properties: The left plot shows the ratio of molecular gas to stellar mass as a function of sSFR. On the right we show the depletion time as a function of sSFR. Further details can be found in Sec.\ \ref{SFR-gas}. The symbols in this plot are same as in Figure.\ \ref{fig:SFR-KS}.}
\label{fig:fgas-tdep}
\end{figure*}

The fraction of gas in the baryonic content of a galaxy, measured by the ratio of gas to stellar mass, is a key factor in understanding the various observed star formation rates. This relative gas content is plotted versus the specific SFR (sSFR=SFR/M$_*$) in the left plot in Figure.\ \ref{fig:fgas-tdep}. The GRG targets follow the main-sequence galaxies of the xCOLD GASS sample, two of which are gas-poor, NGC~315 and NGC~6251.
Together with the gas fraction, SFR depends on the efficiency of star formation with respect to the amount of fuel, that is, SFR/M$_{\rm mol}$, which is the inverse of the depletion time t$\rm _{dep}$. The latter is plotted versus sSFR in the right panel of Figure.\ \ref{fig:fgas-tdep}. Here again, most of our targets follow the MS scaling relations, except for two, NGC~315 and NGC~6251,
which appear to be too efficient to form stars because they lack molecular gas.
Their t$\rm _{dep}$ (the inverse of star formation efficiency) is shorter than expected from the scaling relation.

\begin{table*}[h!]
\caption[]{Other properties of GRGs. We computed P$_{1.4}$, D, and Q$_{\rm Jet}$ from the data available in the literature, whose details can be found in Sec.\ \ref{RPF}. Column 6 indicates whether the GRG shows signs of restarted AGN activity, and column (7) shows their membership in a group (Gr) or cluster (Cl). Sources below the horizontal line are from the literature. The minus indicates non-availability of data or classification.}
\label{tab:pow}
\centering
\begin{tabular}{cccccccc}
\hline
Name & FR   & log P$_{1.4}$& D & HE/LE & Re & Gr/Cl & Q$_{\rm Jet}$\\
&  &   (W Hz$^{-1}$)     & (Mpc) &  &  & & (10$^{43}$ erg s$^{-1}$)\\
(1) & (2) & (3) & (4) & (5) & (6) & (7) & (8) \\
\hline
\\
R1-1 & II  & 24.31   &  1.64   & LE & Yes & No &-\\
R1-2 &  I  & 23.95   &  0.96   & LE &  - & Gr &- \\
R1-3 & II  & 24.71   &  1.05   & HE &  - & Gr & 0.53 \\
R2-1 & II  & 25.20   &  1.52   & HE & Yes &  - & 2.45\\
R2-2 & II  & 24.57   &  1.47   & HE & -  & Gr &-\\
R2-3 &I/II & -       &  2.48   & LE &  - & Gr & 0.36 \\
R2-4 &I/II & 24.41   &  1.96   & SF & No & - & 1.21\\
R2-5 & II  & 25.36   &  2.80   & HE & -  & Gr & 1.77 \\
R2-6 & II  & 25.94   &  1.79   & HE & No  &- & 17.94  \\
R2-7 & II  & 24.71  &  2.24    & LE &  & Gr & - \\
R2-8 & II  & 24.83   &  1.39   & LE & Yes  & GR &- \\
R2-9 & II  & 26.02   &  1.78   & HE & No & - & 16.53 \\
\hline
NGC~315&I& 24.15   &  1.15   & SF & -  & Gr  & 0.16\\
3C~31 &  I  & 24.48   &  0.93   & LE & - & Gr  & 0.54 \\
3C~236& II  & 26.05   &  4.45   & HE & Yes & - & 14.76 \\
3C~326N& II & 25.66   &  2.02   & LE &Yes & - & 10.42\\
\hline
\end{tabular}

\end{table*}

\subsection{Radio power and fueling}
\label{RPF}

The nuclear activity in galaxies is thought to be triggered when the central supermassive
black hole accretes a significant mass of gas from its interstellar reservoir. Typical
accretion phases in AGN may last $\sim$40 million years, with a duty cycle of a few percent
\citep{Shankar2009,Hopkins2010}. 
The GRGs are thought to be at the end of an active and powerful phase of radio 
galaxy evolution. Several hypotheses have been proposed to account for their extreme
sizes, but the role of each factor is still unclear.
The possibilities include the power of the radio jet, implying a higher
velocity \citep{Schoenmakers2000}, the effect of the environment
(jets are able to extend further in voids, but see \citealt{DabhadeSAGAN20}), 
or a restarted activity \citep{Bruni2019, Bruni2020}. In the latter case,
the sources are identified with two jets at different orientations, called double-double,
or X-shaped \citep{Saripalli2013}, and/or a young central source \citep{Machalski2007b}, 
with a flat spectral 
index \citep{Schoenmakers1998}, which has recently been refuelled.

The determination of their gas reservoir can confirm and quantify these fueling mechanisms,
and help to understand the complex processes involved in their radio-jet launching and evolution.
The important clues for the giant nature of GRGs may lie in the accretion rates, which translate into high excitation (HE) or low excitation (LE). The HE objects are defined to have an accretion rate above one-hundredth of the Eddington rate, while the LE objects accrete below this limit. The latter belong to a regime of radiatively inefficient accretion, which foster the formation of radio jets.
  Table.\ \ref{tab:pow} shows our classification of GRGs in our sample with respect to their AGN excitation type, high and low, or HEGRG and LEGRG. We do not find that HEGRGs have higher molecular gas mass than LEGRGs, as was found by \citet{smolcic11} for RGs. \citet{DabhadeSAGAN20} showed that LEGRGs and HEGRGs have distinct properties, such as their respective small size counterparts LERGs and HERGs. We do not have enough statistics here to see a difference with molecular gas fraction.
 
 Recent gas accretion could be violent, through major mergers, or more secular through
 minor mergers of cold gas accretion from cosmic filaments.
 The observation of the molecular component may bring important clues, showing a perturbed
 component in case of galaxy interaction, or a regular disc for secular evolution.
 It has been observed that the detection rate of molecular gas is 
 high in the edge-darkened 
 Fanaroff-Riley (FR) type 1 radio galaxy \citep{Fanaroff1974} and low in edge brightened 
 FR-II type \citep{Evans2005}. More than 90$\%$ of the GRGs have a morphology of FR-II type \citep{DabhadeLoTSS20,DabhadeSAGAN20}, and it is very interesting to determine whether GRGs with FR-II type morphology also have fewer  molecular gas detections.

Table.\ \ref{tab:pow} summarises the FR type, total radio power (P$_{1.4}$), projected linear size (D), excitation state, and jet kinetic power (Q$\rm_{Jet}$) of all GRGs in our sample along with other GRGs taken from the literature. The 1.4 GHz power was estimated using the NRAO VLA Sky Survey \citep{nvss} flux of the GRG at 1.4 GHz. We used the method described by \citet{Qjet_Hardcastle} to estimate Q$\rm _{Jet}$ using the 150 MHz flux from TIFR GMRT Sky Survey \citep{tgss_intema} and other observations from the literature. A similar scheme was used in \citet{DabhadeSAGAN20} to compute the above parameters for a large sample of GRGs.
The excitation state (accretion mode) of GRGs is determined from the mid-infrared fluxes, as in \cite{DabhadeSAGAN20}. It helps separate star-forming objects from objects whose emission is dominated by the AGN, as well as HE from 
LE radio galaxies with the mid-infrared ratios obtained in the WISE bands.
WISE colours have been demonstrated to be very useful diagnostics of the accretion mode of radio galaxies \citep{Gurkan2014}.

\citet{Evans2005} conducted a survey to investigate the molecular content in RGs, where
they found detections in FR-I and compact sources, but rarely in FR-II objects.
It could be expected that restarted RGs, requiring molecular gas to fuel
their activity, would be detected, but only a few have been found to be rich in molecular gas, for example, the post-merger 3C293 \citep{Evans2005,Labiano2014}. \citet{Evans2005} also searched for a possible correlation between the CO luminosity and the 1.4 GHz core luminosity and found none.
We studied the possible effect of the radio jets on the host galaxy
properties by exploring
the connections between Q$\rm_{Jet}$ and parameters derived from our IRAM-30m observations. Based on the current data, we find no correlation between these parameters.
These results may not uncover the whole aspect because our observations with IRAM-30m presents the global cumulative molecular gas properties, and correlations may exist for nuclear region components alone.

GRGs have more intermediate-age stellar populations than the smaller
FR-II radio sources \citep{Kuzmicz2019}.  This is also true for their neighbours within 1.5 Mpc, which show in general younger populations, suggesting triggered star formation due to their environment. This triggering, either through mergers or cold streams, might also be responsible for their giant sizes. A large fraction of GRGs in our sample belong to groups, like R1-2, R1-3, R2-2, R2-3, R2-5, and R2-7 (cf Table.\ \ref{tab:pow}). R2-3, in particular, resides at the centre of a quite
disturbed tight galaxy group. It is one of the galaxies of the UGC9555 triplet, 
hosting a compact flat-spectrum radio source \citep{Clarke2017}. There is
evidence that the AGN has been fueled by recent interactions and minor 
mergers from its group galaxies. None of 
the other galaxies in the sample shows any signs of recent mergers that might be responsible for their fueling. 

Finally, four out of seven of the CO-detected GRGs belong to the restarted class, 
implying that the molecular gas has recently (within $\sim$100~Myr)
refuelled the AGN and radio jets. However, there seems to be a struggle between 
star formation and nuclear activity for the fueling. Some of these galaxies are actively forming stars, and their depletion time is quite short
(740 Myr and 280 Myr for Speca and 3C236, respectively), while others are rather 
quiescent and favour the AGN, like R1-1 and 3C326 (with t$_{\rm dep}$ of
5.5 Gyr and 2.26 Gyr).
Using observations of the Hubble Space Telescope, \citet{tremblay10} have shown bursts of repeated star formation in the host galaxy of 3C~236, which has a dusty disc around the AGN. Because 3C~236 is known to be a restarted AGN, this might be connected with the young starburst property of the host galaxy. 
Based on the studies of GRGs conducted so far and our results, we understand that we observe each GRG in its different evolutionary phase. Therefore we are unlikely to find a similar fueling trend (e.g. cold gas fueling) for all GRGs because the trend closely depends on the AGN accretion state.

In the future, it will be fruitful to observe the GRGs with ALMA to probe the sphere of influence of the black holes that reside in the AGN. This might be a test for the presence of a molecular torus, as was recently found in some low-luminosity AGNs by \citet{combes19}. 

%%%%%%%%%%%%%%%%%%%%%%%%%%%%%%%%%%%%%%%%%%%%%%%%%%%%%%%%%%%%%%%

\section{Summary}
\label{sum}

We have presented our first IRAM results for a sample of 12 GRGs, selected from the GRG catalogue \citep{DabhadeSAGAN20}. We detected the two first rotational lines of CO in one GRG, which is the giant spiral GRG-J2345$-$0449, and report tentative signals for two others (R2-8 and R2-9).
The two-horn shape of the spectra of R1-1 (GRG-J2345-0449) allowed us to determine the concentration of the molecular gas in the giant spiral galaxy. The best-fit model is a roughly exponential disc for the
molecular component, and its half-light radius is comparable to that of the stellar component. 

We added four GRGs from the literature with CO line detections to our sample. Comparing this sample of 16 GRGs with the MS star-forming galaxies at comparable redshift from \cite{Saintonge2017} and a sample of starburst galaxies at similar $z$ \citep{Combes2011}, we find that most (8) of our targets are on the MS, 7 are on the red sequence, and 4 are starbursts. The molecular content of CO-detected GRGs is quite compatible with what is expected from their SFR, although our detected giant spiral appears to be deficient in star formation, with a depletion time of 5.5 Gyr that is about three times longer than the average of the MS. 

Most of the GRGs detected in molecular gas to date are restarted radio galaxies or are in groups, suggesting that the environment plays an important role to refuel the central AGN. The accreted gas also fuels star formation, and the depletion time is sometimes quite short, a few 100 Myr. This is comparable to timescales similar to the spectral ages of the GRGs. Up to now, the number of GRGs with CO detections is low, and more detections are required to draw firmer conclusions.

%%%%%%%%%%%%%%%%%%%%%%%% acknowledgments %%%%%%%%%%%%%%%%%%%%%%%%
\begin{acknowledgements}
We thank the anonymous referee for very constructive comments and suggestions. 
This work is based on observations carried out with the IRAM 30m telescope. IRAM is supported by INSU/CNRS (France), MPG (Germany) and IGN (Spain). PD, FC and JB gratefully acknowledge generous support from the Indo-French centre for the Promotion of Advanced Research (Centre Franco-Indien pour la Promotion de la Recherche Avan\'{c}ee) under programme no. 5204-2. JB, PD and MM would like to thank IUCAA Radio Physics Laboratory and the facilities therein for their support. PD acknowledges the support of Leiden University. This publication has made use of data products from the NASA/IPAC Extragalactic Database (NED). We acknowledge the usage of the HyperLeda database (http://leda.univ-lyon1.fr). The IRAM staff at Pico Veleta are gratefully acknowledged for their help with the 30m observations. This work was supported by the Programme National Cosmology et Galaxies (PNCG) of CNRS/INSU with INP and IN2P3, co-funded by CEA and CNES.
\end{acknowledgements}
%%%%%%%%%%%%%%%%%%%%%%%%%%%%%%%%%%%%%%%%%%%%%%%%%%%%%%%%%%%%%%%

\bibliographystyle{aa}
\bibliography{GRG-IRAM}
%---------------------
\end{document}